\documentclass[sigconf,natbib=true,anonymous=false]{acmart}
\usepackage{caption,subcaption}
\usepackage{hyperref}
\usepackage{amsmath}
\usepackage{natbib}
\usepackage{bm}
\usepackage{cleveref}
\usepackage{upgreek}
\usepackage{mathtools}
\usepackage{nccmath}
\usepackage{url}
\usepackage{amsthm}

\usepackage[inline]{enumitem}
\usepackage{makecell}
\usepackage{boldline}
\usepackage{booktabs}
\usepackage{arydshln}
\usepackage{multirow}
\usepackage[htt]{hyphenat}
\usepackage{soul}
\usepackage{color, colortbl}
\usepackage{xcolor}		

\usepackage{tikz}

\newcommand{\RNum}[1]{\uppercase\expandafter{\romannumeral #1\relax}}

\definecolor{brightmaroon}{rgb}{0.76, 0.23, 0.28}

\definecolor{greeni}{HTML}{9ec5aa}
\definecolor{redi}{HTML}{ec6d57}
\definecolor{whitei}{HTML}{ffffff}
\newcommand{\greenbox}[1]{\colorbox{greeni!42}{#1}}
\newcommand{\redbox}[1]{\colorbox{redi!42}{#1}}
\newcommand{\whitebox}[1]{\colorbox{whitei!42}{#1}}

\copyrightyear{2026}
\acmYear{2026}
\setcopyright{cc}
\setcctype{by}
\acmConference[CHIIR '26]{2026 ACM SIGIR Conference on Human Information Interaction and Retrieval}{March 22--26, 2026}{Seattle, WA, USA}
\acmBooktitle{2026 ACM SIGIR Conference on Human Information Interaction and Retrieval (CHIIR '26), March 22--26, 2026, Seattle, WA, USA}
\acmPrice{}
\acmDOI{10.1145/3786304.3787888}
\acmISBN{979-8-4007-2414-5/2026/03}


\begin{document}
\title{Can Instructed Retrieval Models Really Support Exploration?}

\author{Piyush Maheshwari}
\affiliation{%
 \institution{University of Massachusetts}
 \country{Amherst, USA}
}

\email{psmaheshwari@umass.edu}

\author{Sheshera Mysore}
\affiliation{%
 \institution{Microsoft}
 \country{Seattle, USA}
}
\email{smysore@iesl.cs.umass.edu}

\author{Hamed Zamani}
\affiliation{%
 \institution{University of Massachusetts}
 \country{Amherst, USA}
}
\email{zamani@cs.umass.edu}

\begin{abstract}
Exploratory searches are characterized by under-specified goals and evolving query intents. In such scenarios, retrieval models that can capture user-specified nuances in query intent and adapt results accordingly are desirable — instruction-following retrieval models promise such a capability. In this work, we evaluate instructed retrievers for the prevalent yet under-explored application of aspect-conditional seed-guided exploration using an expert-annotated test collection. We evaluate both recent LLMs fine-tuned for instructed retrieval and general-purpose LLMs prompted for ranking with the highly performant Pairwise Ranking Prompting. We find that the best instructed retrievers improve on ranking relevance compared to instruction-agnostic approaches. However, we also find that instruction following performance, crucial to the user experience of interacting with models, does not mirror ranking relevance improvements and displays insensitivity or counter-intuitive behavior to instructions. Our results indicate that while users may benefit from using current instructed retrievers over instruction-agnostic models, they may not benefit from using them for long-running exploratory sessions requiring greater sensitivity to instructions.
\end{abstract}

\maketitle

\section{Introduction}
\label{sec-intro}
The rise of instruction following capabilities has enabled users to interact with powerful LLMs through natural language instructions \cite{ouyang2022humanprefs}. Following their emergence in generative models, retrieval models that follow user instructions were developed \cite{su2023oneembedder, asai2023task}. These models promise users the ability to specify nuanced relevance criteria, narratives, or aspects with their query. This has extended the frontier of retrieval models from being optimized for semantic similarity to targeting more complex requests.

To enable the development of instructed retrievers, recent work has constructed large-scale benchmarks spanning standard keyword search as well as more complex requests ranging from code or theorem retrieval, tip of tongue retrieval, and literature search, among others \cite{wang2024birco, killingback2025benchmarking, su2025bright}. However, little work has investigated the use of instructed retrievers in the exploratory search scenarios where users start at a seed-document and explore a corpus of documents. Our work bridges this gap through a focused evaluation of instructed retrievers for seed-guided exploration.
\begin{figure*}
\centering
    \includegraphics[width=0.7\textwidth]{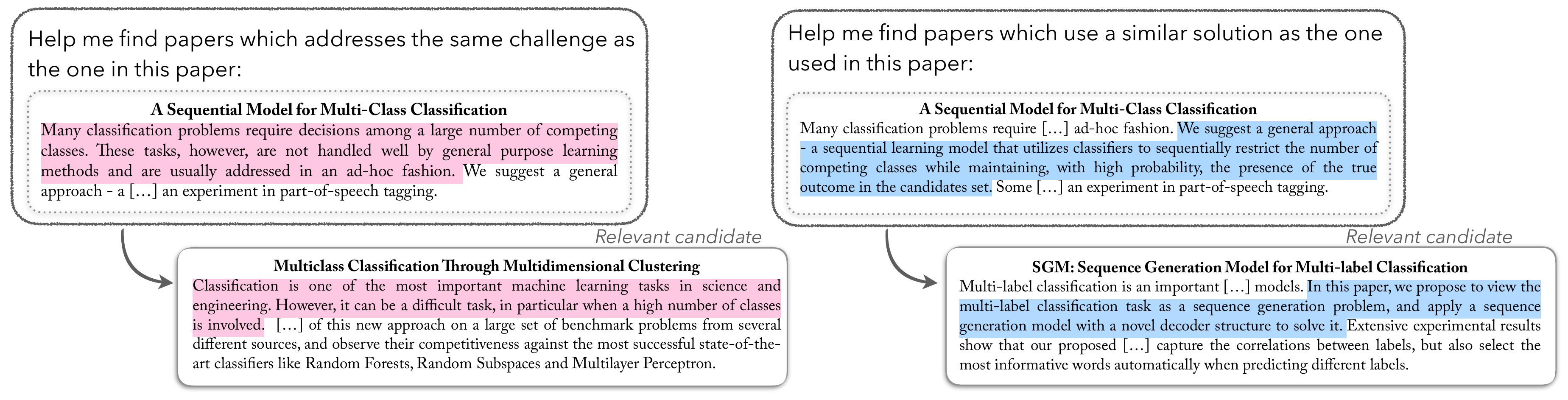}
\caption{CSFCube contains the same query document annotated by experts w.r.t candidate documents for relevance with different instructions. This enables evaluation of both ranking relevance and instruction following for instructed retrievers.}
\label{fig-intro-example}
\end{figure*}

Seed-guided exploration has been the focus of significant research in the design community \cite{ponsard2016paperquest, chau2011apolo} and remains common in exploring document/item collections such as research papers, e-commerce products, and books \cite{zhang2024likethis, mysore2023dslitreviews}. Despite this prevalence, seed-guided exploration remains under-supported in search systems \cite{zhang2024likethis}. The development of highly performing instructed retrievers promises to benefit this prevalent mode of exploration. Specifically, seed-guided exploration is characterized by long document queries where users desire retrieval only based on a few aspects of the document. This calls for retrievers that can perform retrieval based on nuanced aspects of long seed documents \cite{zhang2024likethis, MysoreCSFCUBE2021} -- instructed retrievers offer a promising solution to this challenge. Similarly, the dynamic nature of intents and goals in exploration calls for easily steerable retrieval models \cite{white2009exploratory} -- natural language instructions offer a promising interaction mode for such steerability.

In this paper, we choose the case of scientific document exploration and conduct a careful evaluation of instructed retrievers for their ranking relevances as well as instruction following ability. Retrieval in science offers compelling use cases for accelerating scientific discovery \cite{garikaparthi2025mir, huang2025biomni} and aiding researchers in coping with an exploding literature \cite{chu2021slowedprogress, ajith2024litsearch}. It also enables us to leverage a high-quality, aspect conditional seed-guided exploration test collection \cite[CSFCube]{MysoreCSFCUBE2021} for our experiments. CSFCube consists of expert annotations for paragraph-length queries (paper abstracts) annotated for relevance w.r.t multiple aspects (problem, solution, result) per query. This enables evaluation of the important scenario in seed-guided exploration, where users desire retrieval only based on some aspects of the seed document. Further, CSFCubes' multiple instructions per query (Figure \ref{fig-intro-example}) also enables the evaluation of \emph{instruction following} in instructed retrievers. Instruction following forms a core part of users' experience interacting with instructed retrievers and remains a notably missing aspect in multiple prior benchmarking efforts for instructed retrievers \cite{weller2025followir}. To the best of our knowledge, no other datasets enable both evaluations for exploration applications \cite{do2024multifacet}. In our experiments, we evaluate recent LLMs fine-tuned for instructed retrieval, general-purpose LLMs used with Pairwise Ranking Prompting \cite{qin2024prprankers}, and small parameter LMs finetuned for aspect conditional retrieval.

We find that the best commercial and open-LLM-based instructed retrievers (gpt-4o and GritLM-7B \cite{muennighoff2025generative}) improve significantly over instruction agnostic approaches in ranking relevance. 
However, ranking relevance trends are not reflected in instruction following performance. Models with poor ranking relevance (Llama-3-70B) tended to show better instruction following performance. In experiments examining the sensitivity of the best models to instruction variants, we find that while nuanced instructions improve over generic instructions, extensive tuning of the instructions doesn't deliver improvements. Finally, across experiments, models' sensitivity to instructions remains close to that of instruction-agnostic models, indicating that current models may lack the sensitivity for long-running recall-oriented exploratory sessions. In sum, we contribute the first experimental analysis of instructed retrievers in terms of ranking relevance \emph{and} instruction following for the under-explored application of seed-guided exploratory search. Our work forms an important pre-requisite for human-centered research studying instructed retrievers for exploratory search applications.

\section{Related Work}
\label{sec-related-work}

\textbf{Instruction following in retrieval models.}
The emergence of conversational interfaces for information access has led to an increased focus on building models and systems to support natural language queries \cite{papenmeir2021searchconvs}. Early work in the IR community explored these in the guise of ``verbose'' queries or ``narrative'' queries for applications in web and e-commerce applications \cite{gupta2015verboseq, bogers2017ndr}. 
In recent work, instructed retrievers have unified several forms of complex retrieval interactions spanning constraint-based retrieval, seed-guided retrieval, tip-of-tongue retrieval, and others \cite{killingback2025benchmarking}. Our work extends the understanding of this emerging form of interaction and provides an analysis of instructed retrieval models for aspect conditional seed-guided exploration. 

\textbf{Benchmarking complex retrieval.}
The emergence of complex natural language queries has been accompanied by important benchmarking efforts for such tasks \cite{trippas2024bardanalysis}. \citet[MAIR]{sun2024mair} collate 126 retrieval tasks from multiple domains, which have instructions alongside queries, and benchmark several retrieval models. Our experiments (Section \ref{sec-exp-setup}) use the best performing instructed retriever from their analysis. Going beyond merely including instructions with queries, recent work has focused on curating benchmarks for complex and reasoning-intensive retrieval -- a new frontier unlocked by instructed retrieval. \citet[BRIGHT]{su2025bright} benchmark models on complex requests needing multiple reasoning steps. Closer to our work, \citet[CRUMB]{killingback2025benchmarking} and \citet[BRICO]{wang2024birco} introduce benchmarks for complex retrieval with multi-aspect queries. \citet[INSTRUTIR]{oh2024instructir} and \citet[FollowIR]{weller2025followir} go beyond benchmarking ranking relevance and highlight the importance of evaluating \emph{instruction following}, a core part of user experience with instructed retrievers. Weller et al.\ advocate for having multiple instructions and accompanying relevance judgments per query to evaluate instruction following and introduce the p-MRR metric for instruction following. The multi-aspect relevance judgments of CSFCube (Section \ref{sec-exp-setup}) enable evaluation of this important aspect of instructed retrievers.

\textbf{Retrieval for Science Applications.}
A large body of work has focused on developing retrieval and exploration systems aimed at scientific texts. Early work related to ours focused on citation recommendation \cite{farber2020citation} and analogical search engines to aid scientific brainstorming \cite{kang2022analogyscience}. More recent work has benchmarked LLM-based retrievers for complex literature search queries \cite{ajith2024litsearch} or developed retrieval approaches for RAG systems used for automated scientific discovery \cite{garikaparthi2025mir}. Our work extends these efforts by analyzing instructed retrievers for aspect conditional seed-guided exploration. The closest work to ours is presented by \citet[DORIS-MAE]{wang2023scientific}, who introduce LLM-annotated multi-aspect queries for scientific retrieval. In contrast to \citet{wang2023scientific}, we leverage an expert-annotated test collection enabling greater reliability \cite{deitz2025reliablejudge}. Further, our multi-aspect queries \emph{and} relevances enable evaluation of instruction following (elaborated in Section \ref{sec-exp-setup}). Finally, we contribute an evaluation of multiple families of LLM-based instructed retrievers, going beyond dense retrievers alone.

\section{Experimental Setup}
\label{sec-exp-setup}
\textbf{Test collection.} We use the CSFCube \cite{MysoreCSFCUBE2021} test collection, annotated for aspect conditional seed-guided retrieval by experts in our experiments. It consists of 50 queries and a document collection of 800k documents for retrieval. All queries and documents in CSFCube are drawn from computer science and engineering domains (see examples in Figure \ref{fig-intro-example}). Specifically, CSFCube contains 50 pairs of query aspects and documents. The query aspect indicates which aspect of the query/seed document a retriever should use for retrieval. We format these aspects as instructions for instructed retrievers in our experiments. The aspects focus on  common aspects in most scientific documents \cite{chan2018solvent, kang2022analogyscience} and are diverse enough to generalize to a broader set of user-defined aspects: ``\textit{Background}'', ``\textit{Method}'', ``\textit{Result}''.
Of special note, 32 of the query-aspect pairs in CSFCube annotate the same query document with relevance along two different aspects, which enables CSFCube to be used for instruction following evaluation \cite{weller2025followir}. 
Each query-aspect pair is annotated by at least two experts for relevance against 200 documents on average. This provides a reliable and high-quality test collection to evaluate both ranking relevance and instruction following. 

\textbf{Retrieval Models.} We include a diverse set of models in our evaluation. We evaluate instruction agnostic dense retrievers (\ul{\textsc{Specter2}}, \ul{\textsc{SciNCL}}) specialized for scientific document retrieval \cite{Singh2022specter2, ostendorff2022scincl}. We also include an aspect conditional multi-vector retrieval model (\ul{\textsc{otAspire}}) to evaluate a prior generation of exploratory search model \cite{mysore2022aspire}. Both these model families are built on small parameter models (110M parameters) and form our baseline models. For an embedding-based instructed retriever, we use \ul{GritLM-7B}, the best performing instructed retriever in prior benchmarking efforts \cite{sun2024mair}. Finally, we use a range of open weight (\ul{Mistral-7B}, \ul{Llama-3-8B}, \ul{Llama-3-70B}) and commercial (\ul{gpt-3.5-turbo}, \ul{gpt-4o}) LLMs for ranking with Pairwise Ranking Prompting (PRP) \cite{qin2024prprankers}. 
PRP ranks documents by their win-rate in pairwise relevance comparisons of a pair of candidate documents for a query. It strongly outperforms other prompting methods, such as pointwise scoring, side-steps the long-context limitations of listwise LLM re-ranking, and approaches the performance of trained ranking models. We use an efficient heapsort-based implementation of PRP in our experiments \cite[Sec 3.3]{qin2024prprankers}. We also include a point-wise prompting approach due to its simplicity and wide use. Given the size of the instructed retrievers, we use them as re-rankers to re-rank the first 200 documents from the performant \textsc{SciNCL} first-stage-ranker. We use the expert annotations in CSFCube to report ``\textit{Human}'' performance for ranking relevance and instruction following.

\textbf{Evaluation Metrics.} We evaluate both ranking relevance and instruction following of models. We use NDCG@20 to evaluate ranking relevance and report statistical significance with a paired t-test at $p < 0.05$. We use p-MRR to evaluate instruction following \cite{weller2025followir}.
The metric ranges from $-1$ to $+1$ and indicates worst to best instruction following. p-MRR compares changes in ranking for a single query ($q$) w.r.t two instructions $i$ and $i'$, each with different relevance judgements. The metric is formulated below, where $R$ indicates a document's rank ($1$ is the top rank), and is computed for every document that is relevant for $(i,q)$ but not $(i',q)$:
\begin{equation}
p-MRR =
\begin{cases}
    \frac{R_{i'}}{R_{i}} - 1 & \text{if} ~R_i > R_{i'}\\
    1 - \frac{R_{i'}}{R_{i}} & \text{otherwise}
\end{cases}
\end{equation}
The metric per document is first averaged for a query and then averaged across queries to result in a dataset-level metric. The metric evaluates how well a model follows instruction $i$, with positive scores indicating that the model results in relevant documents ranked at higher positions for $(i, q)$. On the other hand, negative scores indicate that $(i', q)$ results in documents relevant to $(i, q)$ at better ranks, i.e., counterintuitive instruction following behavior. A p-MRR of $0$ indicates that the ranked lists for $(i, q)$ and $(i', q)$ are identical, as expected for an instruction agnostic model. We use the relevance annotations for multiple aspects per query in CSFCube to report p-MRR. For example, we report instruction following for ``\textit{Background}'' treating it as $i$ and \{``\textit{Method}'', ``\textit{Result}''\} as $i'$, etc. Our code details our experiments further: \url{https://github.com/MSheshera/exploration-instructionfollowing}

\section{Results} 
\label{sec-results}
We report ranking relevance and instruction following results in Tables \ref{tab-result-relevance} and \ref{tab-result-instructionf}. In Section \ref{sec-instruction-sensitivity} we probe the sensitivity of instructed retrievers to instruction variants, an important aspect of the user experience interacting with instructed retrievers.

\subsection{Relevance and Instruction Following} 
\textbf{Ranking relevance.} We begin by noting that pointwise (denoted pw) prompting approaches underperform PRP prompting. Similarly, a general-purpose LLM (gpt-4o$_{\text{prp}}$), prompted with PRP, approaches a trained instructed retrieval model (GritLM-7B). These results mirror prior work \cite{qin2024prprankers}. The best performing instructed retrievers (GritLM-7B and gpt-4o$_{\text{prp}}$) improve upon the ranking quality of baseline models. We also see that several open-weight (Mistral, Llama-3) and commercial (gpt-3.5-turbo) models underperform baseline first-stage rankers. In sum, \textbf{our results indicate that the best instructed retrievers meaningfully improve upon baseline models for seed-guided exploration}.
However, we also see that all models lag behind \textit{Human} performance by a large margin, indicating significant room for improvement.

\textbf{Instruction following.} We see that trends in ranking relevance (Table \ref{tab-result-relevance}) are not mirrored in instruction following (Table \ref{tab-result-instructionf}). While Llama-3-70B shows poor ranking relevance, it shows strong instruction following. Similarly, while gpt-4o$_{\text{prp}}$ performs well for ranking relevance, it shows less consistent instruction following performance, varying between different aspects. On the other hand, GritLM-7B balances between ranking relevance and instruction following, displaying more consistent performance on both metrics. We also see \textsc{otAspire}, an aspect conditional retrieval model that uses aspect-specific query sentences to perform retrieval, shows stronger instruction following performance. In Section \ref{sec-instruction-sensitivity}, we see that this strategy also results in improved instruction following for GritLM-7B. As with ranking relevance, we see that \textit{Human} performance leaves a significant margin of improvement for all retrieval models. Further, \textbf{p-MRR scores nearing 0 or negative values indicate that the best instructed retrievers display instruction-agnostic or counter-intuitive instruction following behavior}. This indicates that, despite strong ranking relevance, there remains significant room to improve models for exploratory search applications, which require nuanced forms of relevance and instruction following behavior that is intuitive to users.

\begin{table}[t]
\centering
\caption{Ranking relevance on \textsc{CSFCube}. $^{\times}$ indicates lack of a significant difference from GritLM-7B.}
\scalebox{0.8}{
\begin{tabular}{@{}lcccc@{}} \toprule
Aspects $\rightarrow$ & 
\multicolumn{1}{c}{\textit{Agg.}}   &
\multicolumn{1}{c}{\textit{Background}}       & 
\multicolumn{1}{c}{\textit{Method}}       &
\multicolumn{1}{c}{\textit{Result}}           \\
\cmidrule(lr){2-2} \cmidrule(lr){3-3} \cmidrule(lr){4-4} \cmidrule(lr){5-5}
Models & \small{NDCG@20} & \small{NDCG@20} & \small{NDCG@20} & \small{NDCG@20}\\
\midrule
\emph{Human} & \whitebox{\emph{77.91}} & \whitebox{\emph{82.75}} & \whitebox{\emph{70.38}} & \whitebox{\emph{81.00}}\\
\midrule
\textsc{SciNCL}           & 36.43 & 45.22 & 26.19$^{\times}$ & 38.24$^{\times}$\\
\textsc{Specter2}           & 36.42 & 43.44 & 26.34$^{\times}$ & 40.43$^{\times}$\\
\textsc{otAspire}           & 36.02 & 45.33 & 23.89$^{\times}$ & 39.35\\
\midrule
GritLM-7B           & \whitebox{\textbf{42.38}} & \whitebox{\textbf{53.15}} & \whitebox{\textbf{28.82}} & \whitebox{\textbf{45.90}}\\
\midrule
gpt-3.5-turbo$_{\text{pw}}$    & \whitebox{20.40} & \whitebox{32.16} & \whitebox{09.73} & \whitebox{20.20} \\
gpt-4o$_{\text{pw}}$          & \whitebox{23.71} & \whitebox{41.79} & \whitebox{21.46} & \whitebox{36.87}\\
\midrule
Mistral-7B$_{\text{prp}}$          & \whitebox{27.34} & \whitebox{36.51} & \whitebox{15.58} & \whitebox{30.35}\\
Llama-3-8B$_{\text{prp}}$          & \whitebox{24.40} & \whitebox{18.51} & \whitebox{20.94} & \whitebox{33.48}\\
Llama-3-70B$_{\text{prp}}$          & \whitebox{34.79} & \whitebox{43.50} & \whitebox{24.68$^{\times}$} & \whitebox{36.71}\\
gpt-3.5-turbo$_{\text{prp}}$     & \whitebox{30.82} & \whitebox{37.33} & \whitebox{19.11} & \whitebox{36.40}\\
gpt-4o$_{\text{prp}}$           & \whitebox{\ul{41.07}$^{\times}$} & \ul{52.40}$^{\times}$ & \ul{30.04}$^{\times}$ & \ul{41.43}$^{\times}$ \\
\bottomrule
\end{tabular}
}
\label{tab-result-relevance}
\end{table}

\textbf{Aspect-specific variation.} Finally, we also examine differences across different aspects in Tables \ref{tab-result-relevance} and \ref{tab-result-instructionf}, given that users tend to examine different aspects depending on their application and expertise \cite{ishita2018whichparts}. \textit{Background} instructions see both the best relevance and instruction following, and \textit{Method} sees the worst performance. These results are reflected in multiple studies using the CSFCube test collection \cite{MysoreCSFCUBE2021, do2024multifacet}. \textit{Background} performance is most similar to topical/keyword similarity that models are optimized for and consequently displays strong performance; on the other hand, strong performance on \textit{Method} aspects requires abstract forms of similarity beyond term overlap, a capability most models don't yet possess. 

\subsection{Instruction Sensitivity}
\label{sec-instruction-sensitivity}
Given that instructions serve as the primary mode of user interaction with instructed retrievers, we examine the impact of common instruction/input variants on both NDCG@20 and p-MRR in Table \ref{tab-result-instruction-sensitivity}. We select GritLM-7B given that it balances between strong ranking relevance and instruction following for this analysis. We avoid experimenting with gpt-4o for instruction variants, given its significant expense.
We compare the Base instruction, i.e., short aspect conditional instructions (see Figure \ref{fig-intro-example}) to the following instruction variants:
\begin{enumerate*}
    \item \ul{Generic}: Uses a generic instruction rather than an aspect conditional one, representing a instruction-agnostic approach with instructed retrievers.
    \item \ul{Definition}: Uses a detailed aspect definition resulting in long form instructions. Our definitions are drawn from \citet{MysoreCSFCUBE2021}.
    \item \ul{Paraphrases}: Reports the average metric over three separate paraphrases of the instruction.
    \item \ul{Aspect subset}: Selects the subset of sentences in the query document that correspond to an aspect (available in CSFCube) rather than using the entire aspect. We also use a short aspect-specific instruction. 
\end{enumerate*}

\begin{table}[t]
\centering
\caption{Instruction following results on \textsc{CSFCube} using p-MRR \cite{weller2025followir}. We report scores between $[-1,1]$ scaled by 100.}
\scalebox{0.8}{
\begin{tabular}{@{}lcccc@{}} \toprule
Aspects $\rightarrow$ & 
\multicolumn{1}{c}{\textit{Agg.}}   &
\multicolumn{1}{c}{\textit{Background}}       & 
\multicolumn{1}{c}{\textit{Method}}       &
\multicolumn{1}{c}{\textit{Result}}           \\
\cmidrule(lr){2-2} \cmidrule(lr){3-3} \cmidrule(lr){4-4} \cmidrule(lr){5-5}
Models & \small{p-MRR} & \small{p-MRR} & \small{p-MRR} & \small{p-MRR}\\
\midrule
\emph{Human} & \greenbox{\textit{+25.3}}	 & \greenbox{\textit{+29.8}} 	 & \greenbox{\textit{+22.2}}	 & \greenbox{\textit{+24.0}}\\
\midrule
\textsc{SciNCL}           & \colorbox{gray!10}{0.0} & \colorbox{gray!10}{0.0} & \colorbox{gray!10}{0.0} & \colorbox{gray!10}{0.0}\\
\textsc{Specter2}           & \colorbox{gray!10}{0.0} & \colorbox{gray!10}{0.0} & \colorbox{gray!10}{0.0} & \colorbox{gray!10}{0.0}\\
\textsc{otAspire}           & \greenbox{+3.88} & \greenbox{+4.53} & \greenbox{+3.13} & \greenbox{+3.98}\\
\midrule
GritLM-7B           & \greenbox{+2.02} & \greenbox{+2.99} & \greenbox{+1.14}	& \greenbox{+1.92}\\
\midrule
gpt-3.5-turbo$_{\text{pw}}$    & \greenbox{+0.80} & \greenbox{+2.60} & \redbox{-3.74} & \greenbox{+3.55}\\
gpt-4o$_{\text{pw}}$          & \greenbox{+0.85} & \greenbox{+1.17} & \greenbox{+1.13}	& \greenbox{+0.24}\\
\midrule
Mistral-7B$_{\text{prp}}$          & \redbox{-1.10} & \greenbox{+3.29} & \redbox{-8.67} & \greenbox{+2.07}\\
Llama-3-8B$_{\text{prp}}$          & \redbox{-4.00} & \greenbox{+0.68} & \redbox{-3.06}	& \redbox{-9.59}\\
Llama-3-70B$_{\text{prp}}$          & \greenbox{+4.44} & \greenbox{+8.23} & \greenbox{+0.37} & \greenbox{+4.73}\\
gpt-3.5-turbo$_{\text{prp}}$     & \greenbox{+0.86} & \greenbox{+2.03} & \redbox{-4.20}	& \greenbox{+4.74}\\
gpt-4o$_{\text{prp}}$           & \greenbox{+0.97} & \greenbox{+3.79} & \greenbox{+1.50} & \redbox{-2.39}\\
\bottomrule
\end{tabular}
}
\label{tab-result-instructionf}
\end{table}
\begin{table}[t]
\centering
\caption{Examining the sensitivity of the best performing instructed retriever (GritLM-7B) to instruction variants.}
\scalebox{0.8}{
\begin{tabular}{@{}lcccc@{}} \toprule
Aspects $\rightarrow$ & 
\multicolumn{1}{c}{\textit{Agg.}}   &
\multicolumn{1}{c}{\textit{Background}}       & 
\multicolumn{1}{c}{\textit{Method}}       &
\multicolumn{1}{c}{\textit{Result}}           \\
\cmidrule(lr){2-2} \cmidrule(lr){3-3} \cmidrule(lr){4-4} \cmidrule(lr){5-5}
Models & \small{NDCG@20} & \small{NDCG@20} & \small{NDCG@20} & \small{NDCG@20}\\
\midrule
GritLM-7B (Base)        & 42.38 & 53.15 & 28.82 & 45.90\\
$\hookrightarrow$ Generic & 42.08 & 52.46 & 28.94 & 45.53\\
$\hookrightarrow$ Definition           & 41.43 & 52.54 & 28.18 & 44.25\\
$\hookrightarrow$ Paraphrases          & 42.03 & 52.24 & 28.87 & 45.38\\
$\hookrightarrow$ Aspect subset          & 38.21 & 47.70 & 25.48 & 42.07\\
\midrule
 & \small{p-MRR} & \small{p-MRR} & \small{p-MRR} & \small{p-MRR}\\
\midrule
GritLM-7B (Base)  & \greenbox{+2.02} & \greenbox{+2.99} & \greenbox{+1.14}	& \greenbox{+1.92}\\
$\hookrightarrow$ Generic  & \greenbox{+0.18} & \greenbox{+0.26} & \greenbox{+0.20} & \greenbox{+0.09} \\
$\hookrightarrow$ Definition & \greenbox{+2.01} & \greenbox{+2.38} & \greenbox{+1.22} & \greenbox{+2.44}\\
$\hookrightarrow$ Paraphrases    & \greenbox{+1.53} & \greenbox{+2.70} & \greenbox{+0.95} & \greenbox{+0.93}\\
$\hookrightarrow$ Aspect subset  & \greenbox{+5.89} & \greenbox{+7.60} & \greenbox{+5.96} & \greenbox{+4.10}\\
\bottomrule
\end{tabular}
}
\label{tab-result-instruction-sensitivity}
\end{table}

We note several trends in Table \ref{tab-result-instruction-sensitivity}. We see that Generic instructions result in similar NDCG@20 metrics to the base aspect-specific instruction; however, we also see Generic instructions resulting in p-MRR nearing 0. Indicating that while GritLM-7B may not display strong sensitivity to nuanced instructions, they provide greater steerability than generic and unnuanced instructions. On the other hand, we see that both aspect specific instructions (Definition) or Paraphrases impact NDCG@20 and p-MRR by smaller amounts. This indicates that GritLM-7B remains insensitive to the precise wording of an instruction, likely easing the burden on users to craft the best possible instruction, diverging from interactions with general-purpose LLMs \cite{atreja2025promptscss}. Finally, we see that the Aspect subset strategy delivers the best instruction following performance. 
However, we see this results in worse ranking relevance, likely due to lost query document context. Taken together, these results indicate that the best current instructed retrievers offer meaningful gains in ranking relevance and offer some sensitivity to nuanced instructions, but users may not benefit from extensive instruction tuning and may require more sensitive models to aid long-running recall-oriented exploratory sessions.

\section{Conclusions}
\label{sec-conclusions}
In this paper, we present a focused evaluation of instructed retrievers for the under-explored application of aspect-conditional exploratory search. We make use of an expert-annotated dataset, CSFCube to evaluate both ranking relevance and instruction following in instructed retrievers. Our results indicate that while instructed retrievers improve ranking relevance over instruction-agnostic models, current models show counterintuitive or insensitive behavior to instructions. Instructed retrievers promise an advanced retrieval model for building and studying future human-centered interactive exploratory search systems. However, little work has evaluated human centered aspects of instructed retrievers such as instruction following and sensitivity to instruction variants for exploratory search applications -- our work fills this gap and forms an important empirical basis for future human-centered research.

\begin{acks}
We thank anonymous reviewers for their feedback. This work was supported in part by the Center for Intelligent Information Retrieval. Any opinions, findings and conclusions or recommendations expressed in this material are those of the authors and do not necessarily reflect those of the sponsor.
\end{acks}

\bibliography{facetdiscovery-apps}
\bibliographystyle{ACM-Reference-Format}


\end{document}